\documentclass[twocolumn,prc,showpacs]{revtex4}
\usepackage{graphicx}

\begin{document}
\title{\Large Wobbling excitations and tilted rotation in $^{163}$Lu}
\author{Daniel Almehed}
\email{D.Almehed@umist.ac.uk}
\affiliation{Department of Physics, UMIST, P.O. Box 88, Manchester M60 1QD, 
United Kingdom}
\author{Rashid G. Nazmitdinov}
\affiliation{Departament de F{\'\i}sica, Universitat de les Illes Balears, 
E-07122 Palma de Mallorca, Spain\\
Bogoliubov Laboratory of Theoretical Physics,
Joint Institute for Nuclear Research, 141980 Dubna, Russia }
\author{Friedrich D\"{o}nau}
\affiliation{IKH, FZ Rossendorf, PF 510119, D-01314 Dresden, Germany }

\begin{abstract}
Using a microscopic self-consistent
model, we analyse wobbling excitations
built upon the rotational band in $^{163}$Lu,
which is identified with a rotation of a triaxial, 
strongly deformed shape.
We find that the presence of pairing correlations
substantially affects the energy of the wobbling excitations. 
Our calculations predict an onset of a tilted rotation  
at a critical rotational frequency where the energy 
of the wobbling excitations approaches zero.
\end{abstract}


\maketitle

The study of rotational and vibrational excitations
is a major source of our understanding of 
a nuclear dynamics at low energy.
For instance, the concept of deformation of a nuclear shape, created
by an effective nuclear potential, is well established.
The analysis of interplay of a nuclear shape and orientation 
of an angular momentum led to various important discoveries \cite{BM75,Fr01}.
Recently, rotating non-axially deformed nuclei attract a considerable 
experimental attention. For a long time, experimental data did not provide
an irrefutable proof for existence of the non-axiality. The modern generation of 
detectors opens new avenues to study  nuclear spectra and, in particular, 
low-lying excitations near yrast line with high  precision.
In fact, the non-axial deformation gives rise to a new type of
dynamics involving the orientation degree of freedom. This
includes the chiral rotation \cite{Fr01} and wobbling
excitations \cite{BM75,JM79,Ma79,OH01}.

As a rule, low-lying rotational bands are well described within the cranking 
model with a principal axis rotation. If a nuclear shape becomes non-axial, 
one expects low-lying vibrational excitations built on top of such a rotational
band. These excitations, called the wobbling excitations by analogy 
with a classical wobbling motion, are created by a fluctuation of the angular 
momentum direction around one of the principal axes of a deformed nucleus.
Having in mind that nuclei can rotate also about a tilted axis \cite{Fr01},
one may suggest for non-axial shapes a possible transition from a wobbling mode 
with a rotational axis fluctuating around one principal axis towards a tilted 
rotation  with a fixed rotational axis lying in a plane 
in between two principal axes \cite{HN02}.

We recall that the wobbling excitations in rotating nuclei were suggested 
in Ref. \onlinecite{BM75} and analysed first within 
the microscopic approach in Ref. \onlinecite{MJ78}. 
However, the first experimental evidence of such excitations is reported only 
recently~\cite{OH01,JH02,JH02b,AM03,SH03,GC04,JH04}.

The interest in wobbling excitations was sparked by the discovery of
a particular excited rotational band above the rotational band 
in $^{163}$Lu~\cite{OH01}, which is associated with a triaxial, strongly 
deformed (TSD) nuclear shape. The lowest band is called TSD1 band, while the 
excited one is denoted as TSD2 band.
In accordance with the rates of the observed inter-band $E2$ transitions the
structure of the TSD2 band is identified with wobbling excitations. Later, in 
the same nucleus a second rotational band was found, which was interpreted as a 
two phonon wobbling band~\cite{JH02}. The fact that the two phonon band has less 
then twice the excitation energy of the one phonon band has been seen as a sign 
of non-harmonic vibrations~\cite{JH02,SH03}. The theoretical description
of those band structures was done in terms of the phenomenological, 
particle plus rotor model \cite{Ha02,HH03}. At the same time, 
a non self-consistent, microscopic analysis (within a mean field 
plus random phase approximation approach) were performed in 
Refs. \onlinecite{MS02} and \onlinecite{MS04}. In Ref. \onlinecite{MS04} it was concluded 
that the pairing correlations do not affect the wobbling phonon, and this  
should be considered as a specific feature of such excitations. 
On the other hand, it was also found that the wobbling motion is very sensitive 
to a single-particle alignment. It is well known, however, that the alignment 
decreases the pairing correlations. Therefore, the question arises about the 
validity of this conclusion. One of our goals is to clarify this issue in 
a calculation based on the cranking+random phase approximation 
(called hereafter CRPA), with and without the pairing interaction and 
where we take care of the self-consistency.  
Following the analysis of Ref. \onlinecite{HN02}, we also aim to study a possible 
transition to a tilted mean field solution at the point where the RPA solution 
goes to a zero energy.

We start with a pairing+QQ Hamiltonian~\cite{RS80}:
\begin{equation}
  \label{eq:H1}
  \hat{H} = \sum_k{\epsilon _k} \hat{c}_k^\dagger \hat{c}_{k} -
  \frac{\kappa}{2} \hat{Q} \cdot  \hat{Q}-
  \sum_{\tau=n,p} G_\tau \hat{P}_\tau^\dagger
\hat{P}_\tau.
\end{equation}
Here $\epsilon _k$ are the single-particle 
energies of the spherical modified oscillator 
Hamiltonian $\hat{h}_{\rm sph}$ 
\begin{equation}
  \label{eq:Nilsson1}
  \hat{h}_{\rm sph} = \frac{\hat{p}^2}{2M}+
    \frac{M}{2} \omega_0^2  \hat{r}^2
  - \hbar \omega_0 \tilde{\kappa} \left[ 2 {\bf l} \cdot {\bf s} + \tilde{\mu}
    \left( {\bf l}^2 - \left<{\bf l} ^2 \right>_N \right) \right] .
\end{equation}
The  operators  $\hat{c}^\dagger_k$ ($\hat{c}_k$) are fermion creation
(annihilation) operators with the suffix $k(l)$ labelling a complete set of
quantum numbers. The parameters $\tilde{\kappa}$ and $\tilde{\mu}$ are standard 
ones \cite{NR95}. The quadrupole residual interaction
$\hat{Q} \cdot  \hat{Q} =  \sum_{m=-2}^{2}\hat{Q}_m^2$ is a sum over the
five components of the quadrupole operators built up from $\hat{r}^2
\hat{Y}_{20}$ and the linear combinations $\hat{r}^2 ( \hat{Y}_{2m} \pm
\hat{Y}_{2-m})$ for $m=\pm (1,2)$. The quadrupole operators are defined as 
$\hat{Q}_m = \sum_{kl \tau} q_{m,kl} \hat{c}_{k\tau}^\dagger \hat{c}_{l\tau}$,
$(m=0,\pm 1,\pm 2)$ where $q_{m,kl}\,=\langle k|\hat{Q}_m|l\rangle$ 
and $\tau = \pm 1$ distinguishes  neutrons and protons, respectively. 
The pairing operator $\hat{P}$ has a usual 
form $\hat{P}^\dagger_\tau = 
\sum_{k>0}\hat{c}^\dagger_{k\tau}\hat{c}^\dagger_{\bar{k}\tau}$.
The index $\bar k$ refers to the time conjugated state.

To study rotational properties of the system we perform the Legendre
transformation into the rotating frame
\begin{equation}
  \label{eq:H2}
  \hat{H}' = \hat H -  \vec{\omega} \cdot \vec{\hat{J}}
\end{equation}
where $\vec{\hat{J}}=\left( \hat{J}_x,\hat{J}_y,\hat{J}_z\right)$
is the angular momentum operator and
$\vec{\omega}= \omega \left( \sin \theta \cos \varphi,\sin \theta \sin \varphi,
 \cos \theta \right)$ is the rotational angular frequency vector. 
Here, $\omega$ is the magnitude of the rotational frequency and 
$\theta$ and $\varphi$ are Euler (tilt) angles of the cranking direction.
 It should be pointed out that we are 
not restricted to the signature symmetry used in the approach 
of Refs. \onlinecite{MS02} and \onlinecite{MS04}.

The mean field part of $H'$, Eq.~(\ref{eq:H2}), can be written as
\begin{eqnarray}
  \hat{h}'_{\rm MF} &=& \hat{h}_0  -\vec{\omega}
\cdot \vec{\hat{J}}
  - \frac{2}{3} \hbar \omega_0 \epsilon_2
  \left( \hat{Q}_0 \cos \gamma -\hat{Q}_2 \sin
\gamma \right) \nonumber \\
  \label{eq:H3}
  & & - \sum_\tau \Delta_\tau \left(
\hat{P}^\dagger_\tau + \hat{P}_\tau
  \right) - \lambda_\tau \hat{N}_\tau
\end{eqnarray}
where 
$\Delta_\tau$ are the pair field strengths and $\lambda_\tau$ are 
constraints for
the particle numbers. Further, $\epsilon_2$ and $\gamma$ denote the deformation
parameters in the intrinsic frame of reference defined by the self-consistency conditions:
$ \kappa \left< \hat{Q}_0 \right> = \frac{2}{3}\hbar \omega_0 
\epsilon_2 \cos \gamma$, 
$\kappa \left< \hat{Q}_2 \right> = -\frac{2}{3}\hbar\omega_0 
\epsilon_2 \sin \gamma$,
$\left< \hat{Q}_{\pm 1, -2} \right> = 0$ (see details in Ref. \onlinecite{Fr01}).
Here $<...>$ means the mean field value.
Using the Bogoliubov transformation such as
$\hat{\alpha}_i^\dagger \equiv \sum_k U_{ik} \hat{c}_k^\dagger + 
V_{ik}  \hat{c}_k$, we obtain Hartree-Bogoliubov equations that 
are solved  for the TSD1 band in
$^{163}$Lu, with and without the pairing field. 
The major shells $N=4-6$ are considered for both protons and neutrons. 
We used $\Delta N=0$ quadrupole interaction only. This type of interaction
provides quite reasonable results with regard to equilibrium deformations in the
mean field calculations (see also discussion in Ref. \onlinecite{JPG}). 
Further, this configuration space is sufficient for studying 
the pairing effects on low-lying excitations. 
The pairing strength is adjusted to reproduce the 
global fit of the odd-even mass differences for the ground state of 
even-even nuclei \cite{MN92}. The proton pair field is reduced 
by 20\% due to the odd proton. The quadrupole interaction strength is fitted 
to reproduce the ground state deformation obtained with Nilsson-Strutinsky
calculations. This is done separately with and without pairing. The self-consistent solution 
$\Phi^\omega \equiv |\Omega\rangle$, shortly denoted as $\,\,\rangle$,
corresponds to the minimum of the energy surface
$E'(\epsilon_2,\gamma,\Delta_\tau,\omega,\theta,\varphi)\,=\,\langle\hat
H'\rangle$. This implies that the equilibrium values for all 
these parameters change as a function of the rotational frequency $\omega$.
Note, that our vacuum states $\,\,\rangle$ are 
rotating odd-A particle configurations, similar to the ones used 
in Refs. \onlinecite{MS02} \onlinecite{MS04}.

\begin{figure}[htbp]
  \centerline{\includegraphics[clip,width=7cm]{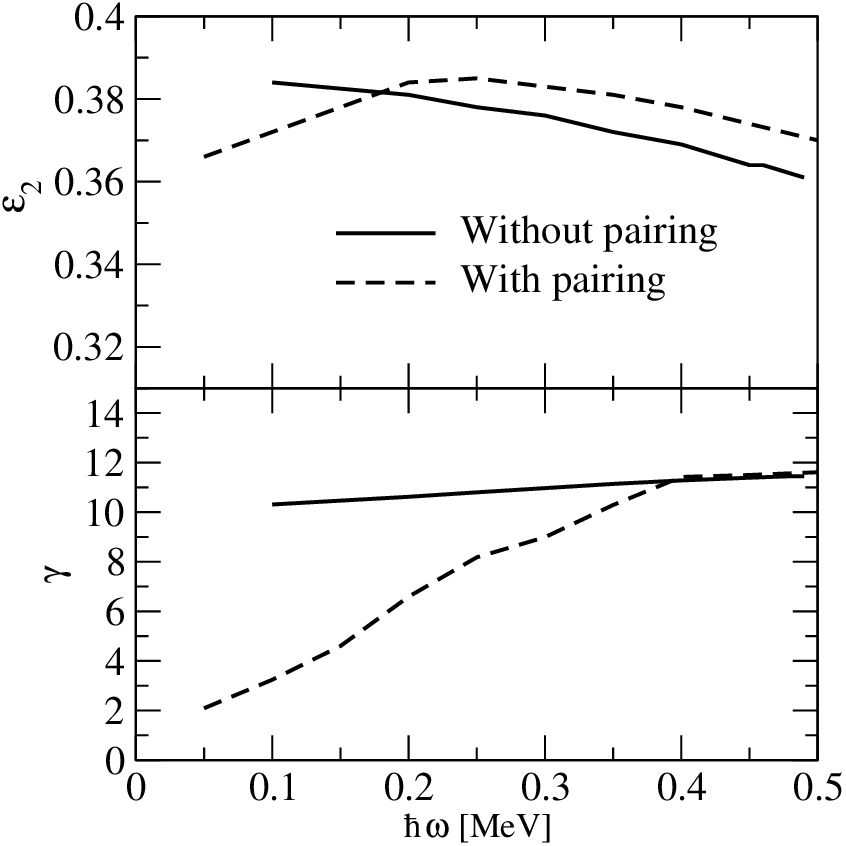}}
  \caption{The deformation parameters $\epsilon_2$ and $\gamma$, associated 
     with the TSD1 band in $^{163}$Lu, as a
    function of rotational frequency $\omega$.}
  \label{fig:Lu163Def}
\end{figure}

For $\hbar \omega < 0.5$ MeV we find a principal axis rotation as the lowest
solution. 
\begin{figure}[htbp]
\centerline{\includegraphics[clip,width=7cm]{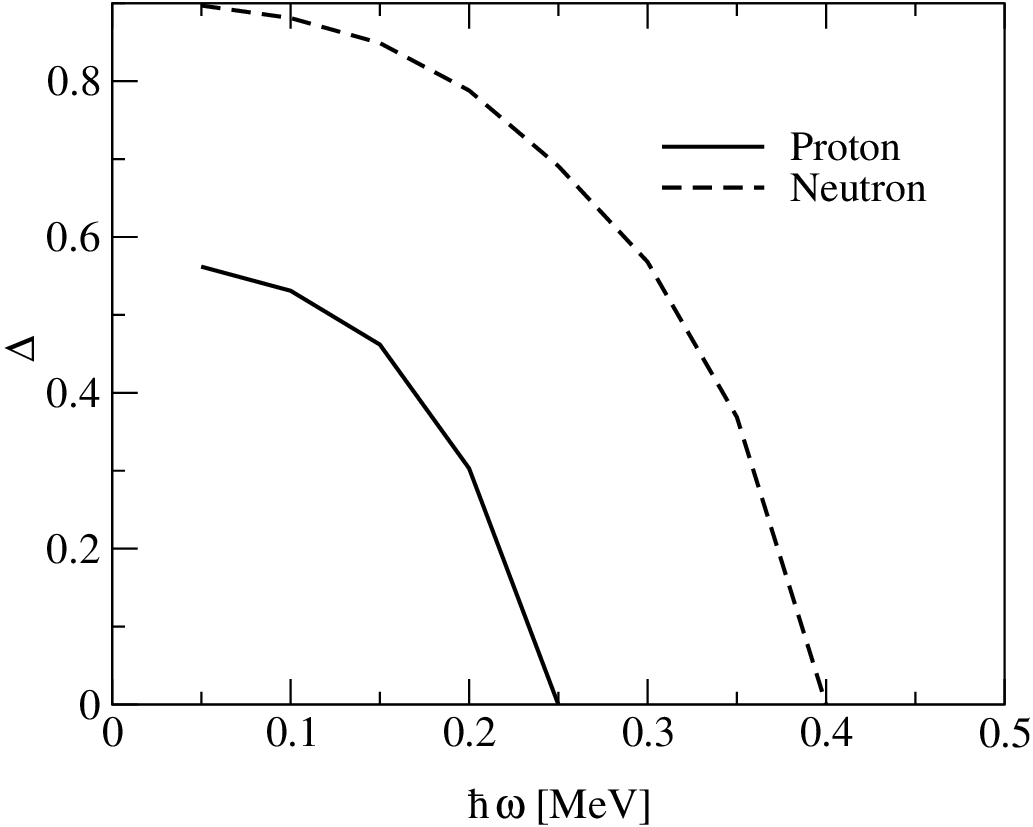}}
  \caption{The neutron and proton pair field in the 
  TSD1 band in $^{163}$Lu as a function of
   the rotational frequency $\omega$.}
  \label{fig:Lu163Del}
\end{figure}
At $\hbar \omega \approx 0.25$ $(0.4)$ MeV  the proton (neutron) pair field
disappears due to the gradual breaking of quasi-particle pairs 
(see Fig.\ref{fig:Lu163Del}).
We also find a increasing $\gamma$-deformation and
a slowly changing $\epsilon_2$-deformation along the band, see
Fig~\ref{fig:Lu163Def}.
At small rotational frequencies strong proton and neutron pair fields 
affect the deformation. In contrast to the unpaired case, the calculations 
with the pairing forces predict a small $\gamma$-deformation 
at low rotational frequencies. Since the triaxial minimum is quite shallow 
in the unpaired calculations, the presence of the pair field is enough 
to almost restore the axial symmetry. With the increase of the rotational 
frequency  the equilibrium deformations manifest a triaxially, strongly 
deformed shape of  $^{163}$Lu.

Once the self-consistent mean cranking solutions are found, we 
apply the quasi-boson approximation in standard way \cite{RS80} 
in order to construct the vibrational wobbling excitations by 
the RPA approach. The particularities of this method for the rotational
case can be found, e.g., in Ref. \onlinecite{KN86}. 
The CRPA Hamiltonian is diagonalised by solving the equations of
motion. As a result, we obtain the determinant of 
the secular equations (see details in Ref. \onlinecite{Al})
which is solved numerically.

The RPA equations have several spurious solutions
related to the symmetries, i.e., the rotational invariance and the particle
number conservation, broken in the mean field calculations. 
If the mean field problem is solved with a high accuracy, the spurious
solutions connected with operators $\hat{J}_z$ and
$\hat{N}_\tau$ will appear at zero energy and the
solution connected with the operator $\hat{\Gamma}^+\sim\hat{J}_x-i\hat{J}_y$
will appear at the rotational frequency. Thus, the spurious solutions 
are completely decoupled from the physical solutions.
In contrast to the approach of 
Ref. \onlinecite{MS04}, we obtain the RPA solutions related to the 
wobbling excitations from the full RPA determinant. 
In Ref. \onlinecite{MS04} this determinant is reduced to a simple dispersion 
equation for the wobbling excitations, which is valid , 
{\it if and only if} all spurious solutions are separated from the physical 
solutions.  Accordingly, the numerical analysis of this dispersion equation alone
is not a warrant for a decoupling of the physical wobbling excitations 
from the spurious solutions. Due to the admixture of the spurious modes 
such an analysis would not be fully reliable.

\begin{figure}%[htbp]
  \centerline{\includegraphics[clip,width=7cm]{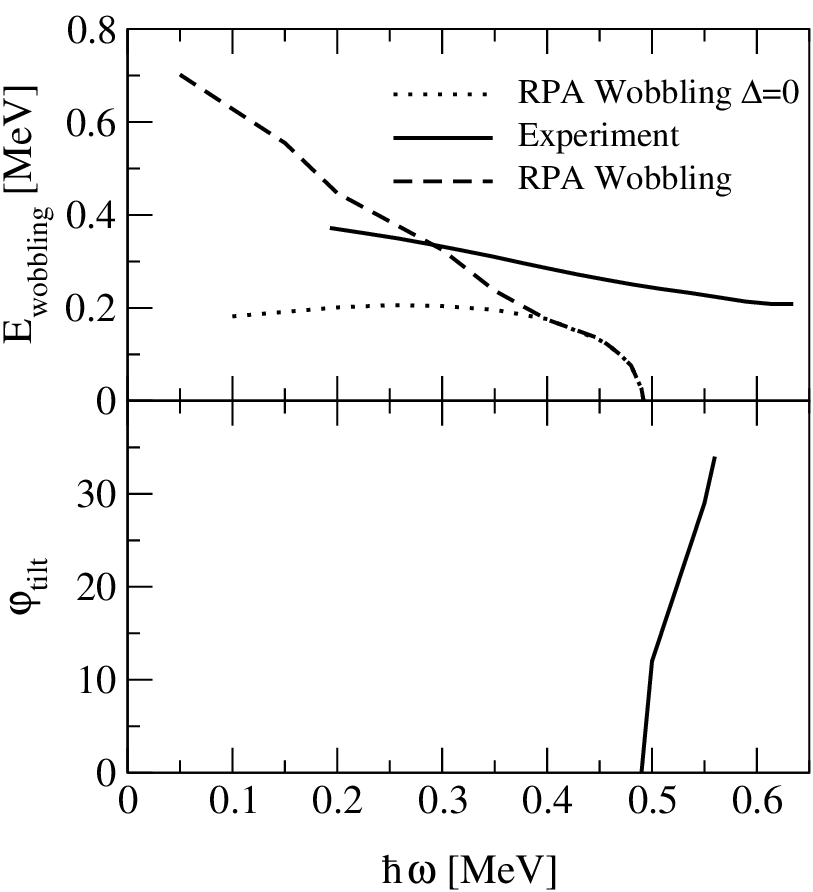}}
  \caption{The wobbling excitation energy  and tilt angle $\varphi (\theta = 0)$ 
    in the TSD1 band in $^{163}$Lu as a function of 
    rotational frequency $\omega$.
    Results with and without pairing correlations are compared 
    to the experimental data taken from~\cite{OH01,JH02}.}
  \label{fig:Lu163WobTilt}
\end{figure}

In Fig.~\ref{fig:Lu163WobTilt} we compare our results of 
the wobbling excitations with experimental data~\cite{OH01,JH02}.
Our results without the pairing correlations are similar to the ones 
obtained in Ref. \onlinecite{MS02}. The calculations predict an almost constant 
wobbling excitation energy up to $\hbar \omega \approx 0.4$ MeV. Above this 
value we obtain a rapid decrease of the wobbling excitations, which leads to 
transition into a stable tilted solution. In contrast to the conclusion of 
Ref. \onlinecite{MS04}, the pairing interaction dramatically changes the results 
at small rotational frequencies. The wobbling excitation energy is substantially 
larger at small rotational frequencies, while it is decreasing with the increase 
of the rotational frequency. The rate of the reduction as a function of the 
rotational frequency is slightly faster than it is seen in the experiment. 
After the collapse of both pair fields at $\hbar \omega \approx 0.4$ MeV, 
the paired and unpaired calculations predict similar results. 
Note, that for zero pair gap even though pairing vibrations are
still present in RPA they do no longer mix with the quadrupole vibrations. 
Large values of the wobbling excitation energy at small rotational 
frequencies are brought about by strong pairing fields 
that reduce $\gamma$-deformation. These results confirm 
the prediction \cite{MJ78} with regard to the $\gamma$ vibrational mode of 
the negative signature. According to Ref. \onlinecite{MJ78}, the increase 
(decrease) of the rotational frequency (the pair field) transforms 
the negative signature $\gamma$ vibrational states with odd spins to
the low-lying wobbling excitations with the increase of the triaxiality.

It is a well known problem of cranked mean field calculations 
that the collapse of the static pair field at high rotational frequency
happens too abrupt. This fact may explain why the 
reduction of the wobbling excitation energy comes too fast 
in our calculations compared to the experimental data. 
As a consequence, we also obtain a transition to a tilted
solution at a relatively low rotational frequency. This
discrepancy could be resolved probably by introducing a self-consistent 
treatment with particle number projection.

\begin{figure}%[htbp]
 \centerline{\includegraphics[clip,width=7cm]{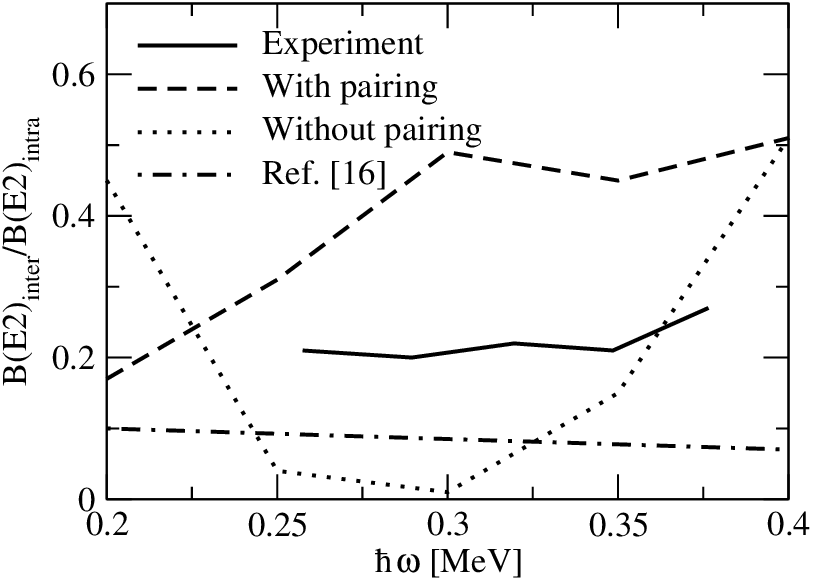}}
  \caption{The ratio of the transition probability from the wobbling band to the
    lower band and the intra-band transitions. 
    The results with and without pairing are compared with the experimental 
     data taken from~\cite{OH01,JH02} and with the results of 
     Ref. \onlinecite{MS02}. We connect experimental 
    points by a solid line for the sake of a comparison.}
  \label{fig:Lu163WobTrans}
\end{figure}

The ratio of the inter-band to intra-band $E2$ transitions for the 
lower part of the wobbling band is shown in Fig.~\ref{fig:Lu163WobTrans}. 
We use an effective charge of $0.5$ ($1.5$) for neutrons (protons) to 
compensate the restricted size of our configuration space. One could observe 
a reasonable agreement with the experimental data, when the pairing correlations 
are included into the calculations, even though the calculated ratio is too large. 
Without the pairing correlations, we obtain a ratio which is smaller than is 
seen in the experiment. For comparison, we included the results from 
Ref. \onlinecite{MS02} which are also smaller than the experimental data.

Summarising, the self-consistent treatment of the wobbling excitations in the
CRPA provides a good description of the experimental data for $^{163}$Lu. 
We conclude that the pairing interactions change the energy of 
the wobbling phonon, at least, indirectly by changing the $\gamma$-deformation.
In addition, our calculations indicate the onset of a tilted rotation in $^{163}$Lu 
above a critical rotational frequency. 
The disappearance of the energy splitting between the  
two signature partner bands is one possible indication for a transition
to a tilted rotational regime. Another indication would be that the two bands are
connected via mixed $M1$ and $E2$ $\Delta I=1$ transitions.  

The authors thank S. Frauendorf for useful discussions.
D.A. acknowledges support from EPSRC (UK). 
This work was partly supported by Grant No.\ BFM2002-03241
from DGI (Spain). R. G. N. gratefully acknowledges support from the
Ram\'on y Cajal programme (Spain).

\end{document}